\documentclass{iau} 
\usepackage{graphicx}

\title[The ESO Spectroscopic facility] 
{ESO Spectroscopic Facility}

\author[Luca Pasquini et al.]   
{Luca Pasquini$^1$,
  B. Delabre$^1$,
   R. S. Ellis $^1$,
  J. Marrero $^1$,
  L. Cavaller $^1$
  \and Tim de Zeeuw $^{1,2}$
}

\affiliation{$^1$ESO, \\ Karl Schwarzschild Str. 2, Garching bei Muenchen, Germany \\ email: {\tt lpasquin@eso.org} \\
$^2$ Sterrenwacht Leiden, Leiden University, \\ Postbus 9513, NL2300RA Leiden The Netherlands \\}

\pubyear{2017}
\volume{334}  
\jname{Rediscovering our Galaxy}
\editors{C. Chiappini, I. Minchev, E. Starkenburg, M. Valentini., eds.}
\begin{document}

\maketitle

\begin{abstract}
We present the concept of a novel facility dedicated to
massively-multiplexed spectroscopy. The telescope has a very wide field Cassegrain
focus optimised for fibre feeding. With a   Field of View (FoV) of
2.5 degrees diameter and a 11.4m pupil, it will be the largest
etendue telescope. The large focal plane can easily  host up
to 16.000 fibres. In addition, a gravity invariant focus for the central
10 arc-minutes is available to host a giant integral field unit (IFU).
The  3 lenses corrector includes an ADC, and  has good performance in the 360-1300 nm wavelength range.
The top level science requirements were developed by a dedicated ESO working
group, and one of the primary cases is high resolution spectroscopy of
GAIA stars and, in general, how our Galaxy formed and evolves. The facility will
therefore  be equipped with both, high and low resolution spectrographs.
We stress the importance of developing the telescope and instrument
designs simultaneously. The most relevant R\&D  aspect is also briefly discussed.
\end{abstract}

\firstsection 
\section{Introduction}

In 2014 ESO carried out a very exhaustive  poll,
that involved several thousand astronomers in the European community (\cite[Primas et al. (2015)]{primas15}),
from which  it clearly emerged  that the most requested facility, not yet available or in construction at ESO,
is a wide field spectroscopic telescope.
As a consequence, ESO set up a scientific working group chaired by one of us (RSE) that delivered a report,
summarising the science cases, setting the main requirements,
and strongly recommending ESO to continue the study of such a  facility (\cite[Ellis et al. (2017)]{ellis17}). 
In parallel, two new concepts for a spectroscopic telescope  were elaborated, one suitable  for fibre feeding, the other for
multi-instruments and mini-IFUs (\cite[Pasquini et al. (2016)]{pasquini16}).

The science cases included studies of the local and high redshift universe, with  a separate aspect
dedicated to time transient spectroscopy and synergies with LSST.
While all readers are invited to read the \cite[Ellis et al. (2017)]{ellis17} report, it is worth mentioning that the
main science case ``the Milky way as a Model Galaxy Organism''  fully embraces the science discussed in this symposium.
It aims at answering fundamental questions such as how
to determine the detailed shape of the galaxy potential and probe the presence
of very low mass halos, to understand how stellar physics exactly works and how the elements are originated,
what is the formation history and memory of a prototypical galaxy, and finally, to compare how the small mass
local group companions fit into the standard model predictions.

Very large surveys, covering a good fraction of GAIA stars (e.g. 85 million stars at limiting magnitude 17),  
have been proposed by the WG to answer to  these questions.

At the same time the quality of the answers  will benefit enormously from retrieving abundances for many elements and with  high precision,
as the power of chemical tagging grows enormously with these two quantities.
Very interesting differences already emerge  amongst
the  Galactic subpopulations when many elements and precise abundances are available
(see e.g. the presentations by R. Smiljanic or  P.E. Nissen at this symposium). High precision,  high resolution spectroscopy is therefore
a strong (and very demanding) requirement. 

The extragalactic science questions center on the evolution of the cosmic web at
high redshift and how galaxies assembled, with emphasis on mapping the redshift
of peak cosmic star formation and beyond.
Several surveys were proposed, and key is the ability to obtain low S/N ratio
spectra for millions of emission line galaxies as faint as I$\sim$
25.8 and high quality spectra  for galaxies at I$\sim$24. 
The added value  of obtaining an integral view of the IGM with a panoramic IFU has been also emphasized. 

The complementarity and follow-up of LSST will open a new domain. It is difficult to predict what the requirements will be,
but it is possible to recognize that, for instance for SNae,  more than 400 events will be present in the FoV at any time, even if  most of them
will be 'expired'. A number of fibres could therefore be dedicated in each field just for this follow-up. 
Most interesting, an independent NOAO study concludes that the most missing critical resource to fully exploit the LSST potential is
to `` Develop or obtain access to a highly multiplexed, wide-field optical multi-object 
spectroscopic capability on an 8m-class telescope, preferably in the Southern Hemisphere'' (\cite[Najita et al. (2016)]{najita16}).

The Australian  Decadal Survey sets a spectroscopic facility at high priority.
The most complete and advanced effort made so-far is constituted by the excellent package developed for the
 Mauna Kea Spectroscopic Explorer (MSE), a proposed telescope for the dome of the CFHT, whose science cases have been compiled by
 \cite[McConnachie et al. (2016)]{mcconnachie16}.

 All these reports testify to the transformational  impact of a 10m-class spectroscopic facility and the vivid interest in the  community worldwide.

 Before moving to a more detailed descripion, it is  worth summarizing the terms of reference:

 - 10m class facility dedicated uniquely to spectroscopic surveys

 - a very large field

 - high multiplex

 - Southern hemisphere location

 Why a facility and not a telescope? Past experience has shown that the well planned survey facilities can survive a
 long time and provide transformational science when they are flexible and conceived as an end-to-end project.
 In the specific case for instance, it is possible to
 optimize the telescope and dome by designing at the same time the instrumentation  needed. In addition, the cost of the
 instrumentation  will be comparable to that of the telescope, so it would be a major mistake not to consider and optimize
 the two together.

 \section{Top Level Requirements}
 
 Here is a summary of the top level requirements (TLRs) provided by  the ESO working group:

 \begin{itemize}

 \item Telescope Parameters: 10-12 meters aperture, versatile (e.g. multi fibre and giant IFU),
   in the Southern hemisphere
 \item Field of View: $\sim$ 5 square degrees
 \item Multiplex: at least 5000 fibres at high spectral resolution
 \item Spectral Resolution: R=20-40000 for High Res, R=1000-3000 for Low Res
 \item Wavelength range: 360-1000 nm
 \item Panoramic IFU: 3x3 arcmin FoV, with R$\sim$5000 and coverage in the blue down to 325 nm

 \end{itemize}

 In  the current design phase, some additional requirements were added to the working group TLRs:

 \begin{itemize}
 \item Optimize costs: Use existing (ESO) Observatory infrastructures, use, whenever applicable, ELT technology and components (e.g. M1 segments)
 \item Enhance present capabilities: The spectroscopic facility shall outperform in survey power the presently
   planned facilities by at least one order of magnitude.
 \item The low resolution spectrographs shall be  usable by the fibres and by the panoramic IFU
 \item Simultaneous observations: it shall be possible to acquire simultaneously low and high resolution observations over the whole FoV
 \end{itemize}

 \section{A powerful telescope } 

 The Pasquini et al. (2016) paper considered two potential designs. The requirement to access the largest possible
 focal plane with fibres, has driven to furher develop  the Cassegrain fibre concept, shown in Figure 1.

 The telescope pupil  is 11.4 meters in diameter. The primary consists of 78 ELT primary segments,
 and the secondary is 4.2m in diameter,
 the same size of the ELT secondary.
 The three lens corrector (1.8 m diameter largest lens) works also as ADC and provides a corrected 2.5 degrees, 
 1.43m diameter focal plane at F/2.86, an excellent aperture for fibre coupling. 
  By inserting two mirrors in front of the corrector, the central 10 arcminutes  of the FoV can be directed to a
 coud\`e gravity invariant focus that can host the low resolution spectrographs and the panoramic IFU, with excellent image quality.

 This telescope is very compact, it does not need  extended Nasmyth platfoms, so the dome size can be kept small when
 compared to most telescopes of similar size.
 A concept of the mechanical structure is given in Figure 2.
 The first floor below the telescope hosts the high resolution
 spectrographs, sitting in a comfortable gravity invariant room, while the room below, with a rotating floor to
 compensate for the field rotation in the IFU, hosts the low resolution ones. 
 
 The combination of large FOV and effective diameter results in a telescope  with the largest
 etendue ever,
 as shown in  Table 1. Its etendue is even larger than LSST because,
 in spite of the huge LSST FOV, its effective area is only $\sim$40$\%$  of  our concept.

\begin{table}
  \begin{center}
  \caption{Overview of etendue for several survey telescopes.}
  \label{tab1}
 {\scriptsize
  \begin{tabular}{|l|c|c|c|c|c|}\hline 
{\bf Name } & {\bf Tel. Dia.} & {\bf Central Obs.} & {\bf Surface (m$^2$)} & {\bf $\Omega (deg^2)$} 
   &  {\bf Etendue} \\ \hline
VLT VIMOS  & 8.0  & 0.97 & 48.75 & 0.043 & 2.08 \\ 
VLT Flames & 8.0  & 0.97 & 48.75 & 0.136 & 6.63 \\
VISTA 4MOST & 3.7 & 0.89 & 9.57  & 4.00  & 38.3 \\
VLT MOONS   & 8.0  & 0.97 & 48.75 & 0.136 & 6.63 \\
WEAVE      & 4.2  & 0.88  & 12.2  & 3.14  & 38.3 \\
SUBARU PFS & 8.0  & 0.97 & 48.75  & 1.33  & 64.7 \\
MAYALL DESI & 3.8 &  0.85 & 9.6    &  8    &  77  \\
LSST (Imaging) & 8.2 & 0.63 & 33.27 & 9.62 & 320 \\
MSE           & 11.2 & 0.97 & 96.0 & 1.50  & 144 \\
ESO Concept   & 11.4 & 0.86 & 87.89 & 4.91 & 431 \\ \hline
  \end{tabular}
  }
 \end{center}
\end{table}

\begin{figure}[h]
\vspace*{-2.0 cm}
\begin{center}
\includegraphics[angle=270, width=6.4in]{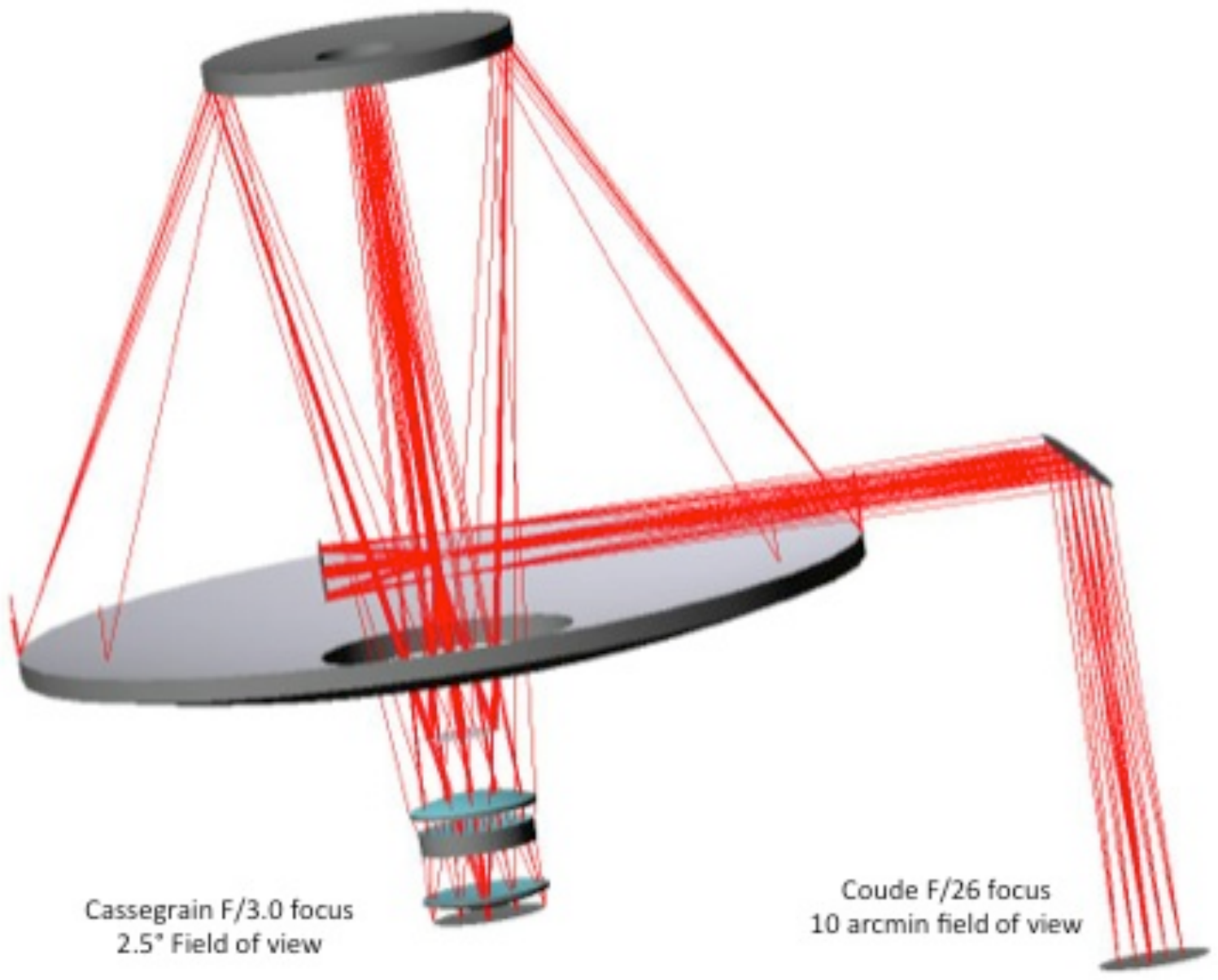} 
\vspace*{-1.0 cm}
\caption{Optical concept of the telescope, with Cassegrain fibre focus and coud\`e for panoramic IFU.}
\label{fig1}
\end{center}
\end{figure}

 \begin{figure}[h]
\begin{center}
 \includegraphics[width=5.4in]{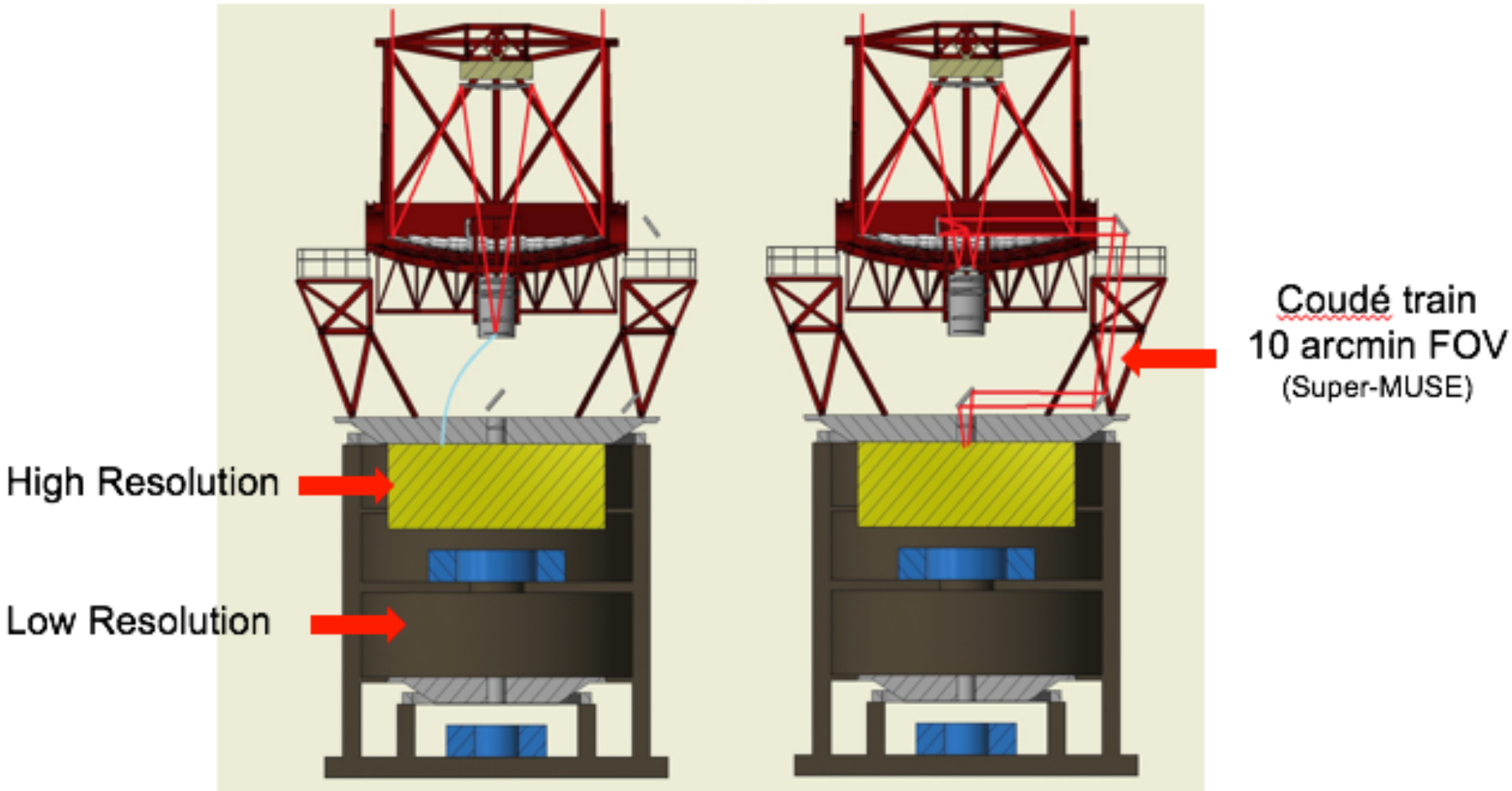} 
 \caption{Telescope with  main structure.}
   \label{fig2}
\end{center}
\end{figure}

 \subsection{The Corrector \& Positioner}

 The working principle of the corrector is inspired by those proposed for NTT, VISTA \& MSE
 (\cite[Grupp et al. (2012)]{grupp12}, \cite[Gillingham \& Saunders (2014)]{gilli14}, \cite[Saunders \& Gillingham (2016)]{saunders16}); 
 the last two lenses and the focal plane  tilt around a circle of 5m radius to
 compensate for the changing atmospheric
 dispersion, as described in Figure 3. The tilt introduces coma, that is corrected by simultaneously
 translating the secondary. The correction is good up to airmass $\sim$1.8,
 and in the 360-1300 nm interval.
 The corrector has only three aspheric surfaces (each lens has one aspheric surface). The glass is silica.  
 The 1.8m aspheric lenses would  probably be the largest ever produced,
 larger, but comparable to the LSST corrector ones. 

 The  large corrected focal plane, combined with a  $\sim$ 160 micron/arcsecond plate scale,
 is very comfortable for installing fibre positioners.
 For instance, by using the positioner mechanism adopted by the DESI survey,
 more than 15000 positioners can be hosted, just using existing technology.
 By adopting a slightly different arm design,
 like the one chosen by MOONS,
 each point of the focal plane can be reached by three fibres.
 This would allow a very convenient ratio of 2:1 between the High resolution and the Low resolution fibres, and
 each point of the FOV is reachable by at least one high resolution fibre and two low resolution ones. With a reference number of 15000 fibres,
5000 objects could be observed at high resolution and 10000 at low resolution simultaneously.

With such a huge multiplex and  etendue,  this facility will have a survey power (etendue$\times$ number of objects)
more than 10 times higher than any other spectroscopic facility either in construction or in design.

 \begin{figure}[h]
\begin{center}
 \includegraphics[angle=270,width=4.4in]{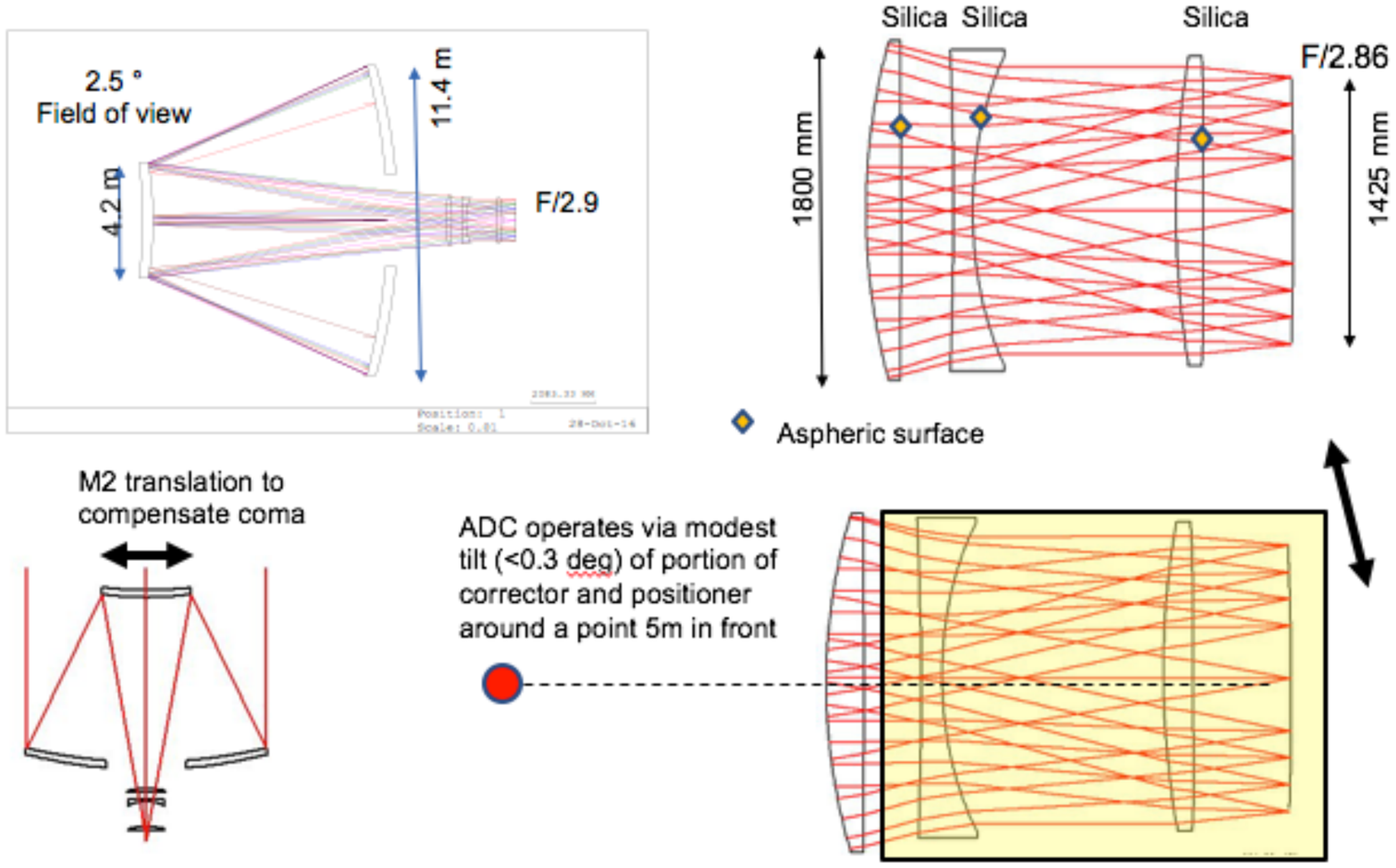} 
 \caption{The Cassegrain corrector design and ADC concept. The last two lenses and the focal plane tilt around a point 5m away to compensate for atmospheric dispersion. The coma introduced is compensated by translating the secondary. }
   \label{fig2}
\end{center}
\end{figure}

\section{The spectrographs }

So many objects, on the other hand, will require  a large number of spectrographs. 
The challenges in building the spectrographs for this facility are to find a reasonable design that can contain the costs.
In order to fully exploit the telescope diameter, in fact,
it is needed to produce quite fast cameras (around F/1) for pixel matching. Another challenge
is to find a suitable design for the  high resolution spectrographs,
which will need a large pupil in order to obtain the required resolution.

In order to procure efficient,  fast and cheap cameras, suitable for multi-spectrographs solution, we
think that it is imperative to use curved detectors. Curved CCDs have been  produced and tested in the past
(\cite[Iwert et al. 2012]{iwert12})
and a lot of effort is presently ongoing on CMOS devices with very good results
(\cite[Hugot et al. (2016)]{hugot16}, \cite[Guenter et al. 2017]{guenter17})
so having them available for this  facility carries a very limited risk.
Curved detectors allow very simple, therefore efficient, cameras, that do not require exotic glasses, therefore cheap. 
The optical design of a low resolution spectrograph is shown in Figure 4 and the F/1.1 cameras are made by just 4
lenses glued in two groups.
The spectrograph has two arms (Blue and Red),
with a separation around 680 nm, and each uses a 4K$\times$4K detector.
With an F/1.1 camera, the scale is of 0.25 arcseconds/pixel and each spectrograph can host $\sim$ 600 one arcsecond
fibres. Less than 20 such spectrographs will suffice to cover all low-res
fibres, while 65 of them will be needed to provide a 3x3 arcminutes panoramic IFU.

As for the high resolution spectrographs, their design is pending, likely
the principle will be similar to those developed for 4MOST, with the novelty  of
using also for these spectrographs the fast cameras allowed by the curved detectors.

 \begin{figure}[h]
\begin{center}
 \includegraphics[angle=270, width=4.4in]{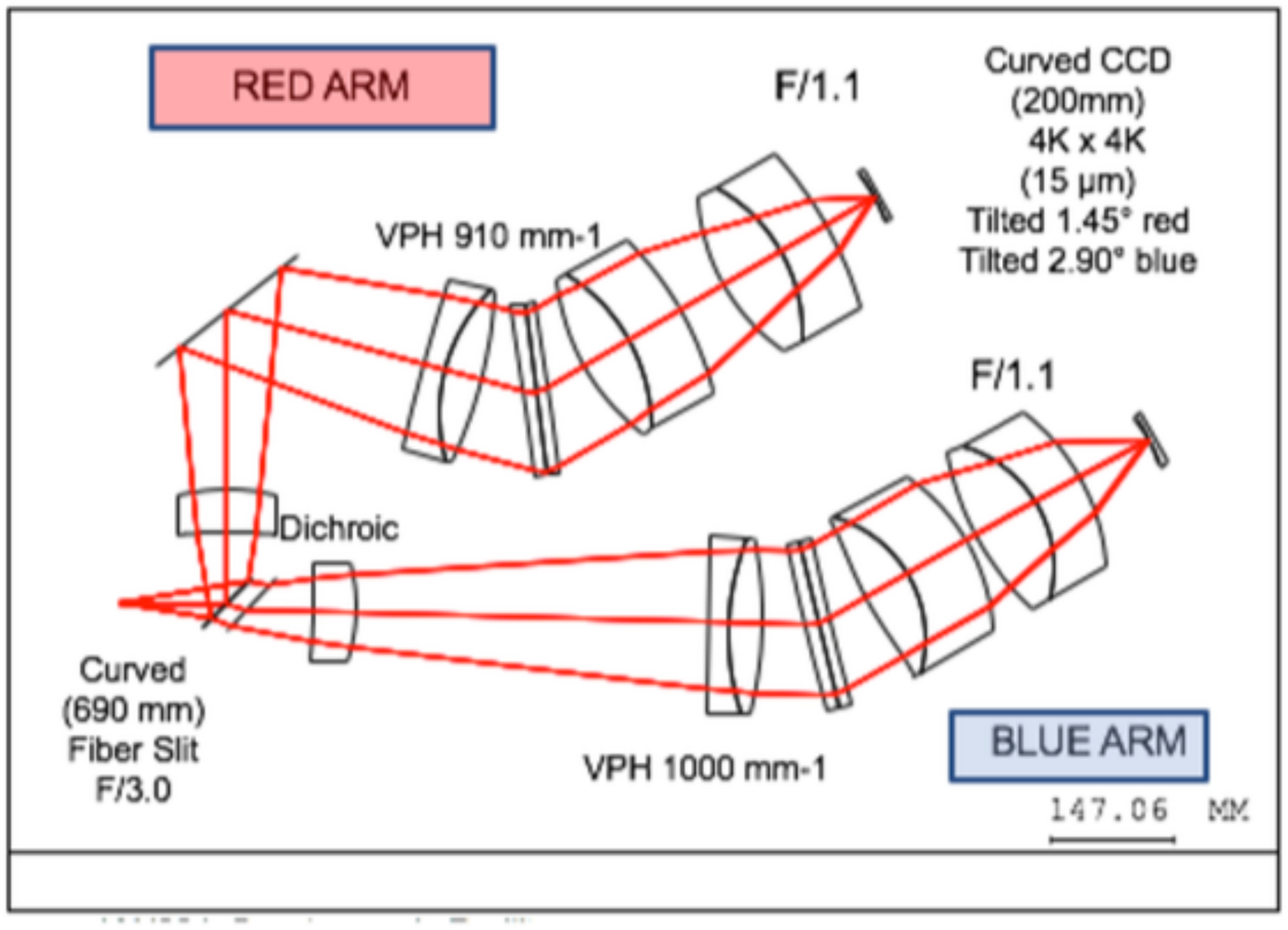} 
 \caption{ Optical design of the low resolution spectrographs. The two arm design  covers the 380-100 nm range. It provides a resolving power of $\sim$2600 with 1 arcsecond fibres, and the double for the panoramic IFU. It accomodates $\sim$ 600 spectra on a 4Kx4K curved detector. }
   \label{fig2}
\end{center}
\end{figure}

\end{document}